\documentclass[11pt]{article}

\usepackage{latexsym}
\usepackage[dvips]{graphicx,color}
\usepackage{graphics}
\usepackage{amsmath,amssymb}
\usepackage{float}
\usepackage{multicol}

\newcommand{\beq}{\begin{equation}}
\newcommand{\feq}{\end{equation}}

\title{\bf Non-Markovian dynamics and von Neumann entropy evolution of a qubit in a spin environment}
\date{}
\author{J. Rodr\'iguez Garz\'on$^{1,2}$\footnote{Electronic Address: jaime.rodriguez@uan.edu.co} and R. M. Guti\'errez$^{2}$\footnote{Electronic Address: rafael.gutierrez@uan.edu.co} \\$^{1}$\emph{Departamento de f\'isica, Universidad Antonio Nari\~no}\\ $^{2}$\emph{Centro de Investigaciones, Universidad Antonio Nari\~no}\\ \emph{Carrera 3 Este No. 47A - 15 Bloque 4, piso 4}\\
\emph{Bogot\'a, Colombia}}

\begin{document}

\maketitle

\begin{abstract}
The dynamics of a central spin-$1/2$ in presence of a local magnetic field and a bath of N spin-$1/2$ particles is  studied in the thermodynamic limit. The interaction between the spins is Heisenberg XY type and the bath is
considered to be a perfect thermal reservoir. In this case, the evolution
of the populations of the reduced density matrix are obtained for different temperatures. A Born approximation is made but not a Markov approximation resulting a non-Markovian dynamics. The measure of the way that the system mixes is obtained by means of the von Neumann entropy. For low temperatures, results show that there are oscillations of populations and of the von Neumann entropy, indicating that the central spin becomes a pure state with \emph{characteristic} time periods in which it is possible to \emph{extract} or recuperate information. In the  regime of high temperatures, the evolution shows a final maximum mixed state with entropy $S=\ln 2$ as it is expected for a two level system.
\end{abstract}

\section{Introduction}
\noindent
The most promising systems that could be scaled to make practical realizations of quantum computation and quantum information are the spin systems in solid state nanostructures \cite{dsarma,friesen, awschalom, DiVincenzo1}. Developments in the last few recent years, permit to manipulate, tune and have full control over \emph{the spin} of individual defined number of electrons in semiconductor nanostructures like quantum dots\footnote{It is important to note that measurements and control over the electron charge is possible, almost, since 1990s, but not so over the spin.} \cite{gammon, kouwenhoven2001}. This particular experimental abilities of control over the spin had motivated the study of individual spins as the fundamental system \cite{kouwenhoven}. However, the correlations of the spin with many degrees of freedom of the surrounding environment lead to finite lifetime of quantum superpositions, bringing pure states into mixed ones.\\
The principal source of noise in this solid state spin nanodevices arises from the hyperfine interaction with nuclear spins; a lot of efforts has been devoted to model spin bath systems \cite{khaetskii3, sousa, schliemann}. Different theoretical proposals have been made in order to model these systems, however, a special interest have been devoted to the configuration known as \emph{spin star network} \cite{hutton, breuer}.  This configuration consists of a $N+1$ spin-$1/2$ particles, where the system of interest is a central spin and the remaining $N$ spins surround the central spin at equal distances. This configuration implies a spatial symmetry which allows a formal and exact analytical solution. This N surrounding spins act as a thermal bath.\\
In the same way, different theoretical calculations have been made in order to obtain the reduced dynamics of the central spin. Many of these works have considered Ising type interaction \cite{luca, san}. The work of reference \cite{hutton}, consider a star network where the interaction between the  spins of the bath is neglected. A similar situation is considered in reference \cite{breuer}, where the interaction between the central spin and the bath is Heisenberg XY type and the the bath is taken in an unpolarized infinite temperature state which in practice is not a feasible realization because of the Coulomb blockade \cite{kouwenhoven}. In the reference \cite{xiao} it is considered a more general case of the dynamics of two central interacting spins in a bath of interacting spins in order to describe the entanglement between the two central spins without considering the entanglement with the bath.\\
In this paper, we consider the reduced dynamics of one central spin in a bath of interacting spins in a configuration of spin star network. The central spin interacts via Zeeman effect with a local magnetic field, the interaction between spins is Heisenberg XY type and the thermodynamic limit is considered. The methodology is similar to that of \cite{xiao}, however, we consider the dynamics of one single spin and the way that it becomes mixed. In addition, in reference \cite{xiao} it is considered a spin wave theory by means of the Holstein-Primakoff transformation \cite{primakoff} and we study the problem of small fluctuations in the context of  Schwinger bosons \cite{schwinger}. The \emph{Schwinger transformation} and the Holstein-Primakoff are closely related as it is shown, however, Schwinger bosons have more natural physical interpretation.
The principal reason to consider the system of one central spin is that they are really controllable systems \cite{gammon, kouwenhoven2001, kouwenhoven} and fulfill the criteria of DiVincenzo for universal quantum computing \cite{DiVincenzo1, DiVincenzo2}.\\
The paper is organized as follows. In Section \ref{sec:model} we present the model by means of the Schwinger's oscillator model of angular momentum \cite{schwinger} and in the thermodynamic limit we obtain a Jaynes-Cummings Hamiltonian type. Then the reduced dynamics of the central spin and the evolution of the populations of the central spin is obtained.
In Section \ref{sec:entropy} we calculate the time evolution of the von Newmann entropy. Different temperature regimes are considered showing, how the temperature influences the mixing of the state. Conclusions are drawn in Section \ref{sec:conclusion}.

\section{The model}\label{sec:model}
\noindent
The characteristics of the model are similar to that of \cite{xiao}, but our principal interest is to describe the coupling  of the central spin with the environment.
We consider a spin star network of $N+1$ interacting spin-1/2 particles. The interaction is Heisenberg XY type and the system of interest is the central spin in presence of a local magnetic field. The bath is represented by the $N$ 1/2 spin particles surrounding the central spin. The total Hamiltonian is
\begin{equation}\label{eq:hamilton}
H=H_S+H_B+H_{SB},
\end{equation}
where $H_S$ and $H_B$ are the Hamiltonians of the system and the bath respectively. The interac\-tion Hamiltonian between the system and the bath is denoted by $H_{SB}$. Reminding that there is a local magnetic field interacting only with the central spin, and that all particles interact via Heisenberg XY, each of this terms can be written as \cite{xiao}
\begin{equation}
H_{S}=\mu_{0}S^z,
\end{equation}
\begin{equation}
H_{B}=\frac{g}{N}\sum_{i\neq j}^N\, S_i^+S_j^- + S_i^-S_j^+,
\end{equation}
and
\begin{equation}
H_{SB}=\frac{g_0}{\sqrt{N}}\left[S^+\sum_{i=1}^N S_i^- +  S^-\sum_{i=1}^N S_i^+  \right],
\end{equation}
where, $\mu_0$ represents the local magnetic field in the $z$ direction, $g$ is the coupling constant between the spins of the bath, and $g_0$ is the coupling constant between the central spin and the bath spin particles. We should stress that the coupling constants  has been scaled as $g/N$ and $g_0/\sqrt{N}$ in
order to obtain the thermodynamic limit \cite{xiao, frasca}. 	
The correspondig \emph{raising} and \emph{lowering} spin operators, written in terms of the usual spin operators $S^x$, $S^y$, are $S^\pm=S^x \pm i S^y$.
The total spin of the bath can be written as $J_{\pm}=\sum_{i=1}^N S_i^{\pm}$, so that the central spin couples to an \emph{effective} collective bath of angular momentum $J$. In terms of $J_{\pm}$ the bath and the interaction terms of the Hamiltonian become:
\begin{equation}
H_{B}=\frac{g}{N} \left(J_+J_- + J_-J_+\right)-g
\end{equation}
and
\begin{equation}\label{sb}
H_{SB}=\frac{g_0}{\sqrt{N}}\left(S^+J_- + S^-J_+
\right)
\end{equation}
respectively.
The spin-$\frac{1}{2}$ operators can be mapped to Bose type operators by means of the connection between the algebra of the angular momentum and the algebra of two independent oscillators. The problem has been studied by J. Schwinger and the process corresponds to
spin wave theory in order to describe the spins in terms of small fluctuactions \cite{schwinger}.

\subsection{Schwinger's bosons}
Consider two non-coupled harmonic oscillators, or \emph{Schwinger  bosons}, denoted by $b$ and $a$. The usual number operators are given by
\begin{equation}
N_b\equiv b^{\dagger} b, \ \ \ \ \   N_a\equiv a^{\dagger} a.
\end{equation}
It is considered that $b$ and $a$ are independent and follow the Bose statistics:
\begin{equation}
[b,b^{\dagger}]=1 \ \ \ \ \ \ [a,a^{\dagger}]=1,
\end{equation}
\begin{equation}\label{noint}
[b,a^{\dagger}]=0 \ \ \ \ \ \ [a,b^{\dagger}]=0.
\end{equation}
Then $N_b$ and $N_a$ conmute, $[N_b,N_a]=0$, and there exist simultaneous eigenkets $\mid n_b,n_a\rangle$ where the correspondig actions of the $b$ and $a$ operators are as ussually:
\begin{equation}
b^{\dagger}\mid n_b,n_a\rangle=\sqrt{n_b+1}\mid n_b+1,n_a\rangle, \ \ \ \ \
a^{\dagger}\mid n_b,n_a\rangle=\sqrt{n_a+1}\mid n_b,n_a+1\rangle
\end{equation}
and
\begin{equation}
b\mid n_b,n_a\rangle=\sqrt{n_b}\mid n_b-1,n_a\rangle, \ \ \ \ \
a\mid n_b,n_a\rangle=\sqrt{n_a}\mid n_b,n_a-1\rangle.
\end{equation}
By successive applications of the creation operators $b^{\dagger}$ and $a^{\dagger}$ we can obtain the most general eigenkets,
\begin{equation}
\mid n_b,n_a\rangle=\frac{(b^{\dagger})^n (a^{\dagger})^n}{\sqrt{n_b!}\sqrt{n_a!}}\mid 0, 0\rangle.
\end{equation}
The \emph{Schwinger transformation} is defined as (with $\hbar=1$)
\begin{equation}
J_+\equiv b^{\dagger}a \ \ \ \ \ J_-\equiv a^{\dagger}b.
\end{equation}
and
\begin{equation}
J_z=\frac{1}{2}(N_b-N_a).
\end{equation}
It is easy to verify that while $b$ and $a$ follows a Bose statistics, the $J_{\pm}$ satisfy the usual relations of
commutation of angular momentum,
\begin{equation}
[J_+,J_-]=2J_z,
\end{equation}
\begin{equation}
[J_z,J_{\pm}]=\pm J_{\pm}
\end{equation}
and acting on the common base $\mid n_b, n_a\rangle$ they give
\begin{equation}
J_+ \mid n_b, n_a\rangle=\sqrt{(n_b+1) n_a} \mid n_b+1, n_a-1\rangle,
\end{equation}
\begin{equation}
J_- \mid n_b, n_a\rangle=\sqrt{n_b(n_a+1)}\mid n_b-1, n_a\rangle,
\end{equation}
\begin{equation}\label{jzeta}
J_z \mid n_b, n_a\rangle=\frac{1}{2}(n_b-n_a)\mid n_b, n_a\rangle.	
\end{equation}
The Schwinger transformation has an immediate physical interpretation. There are two kinds of collective excitations, and  equation (\ref{jzeta}) suggests that $n_b$ is the number of spins up, while $n_a$ is the number of spins down. So, the two kind of collective excitations are: $b^{\dagger}$ creating units of spin up and $a^{\dagger}$ creating units of spin down. The meaning of $J_+$ is that it destroys one unit of spin down at the same time that it creates a unit of spin up. Similarly, $J_-$ destroys a unit of spin up and creates one of spin down. The different processes occur without interactions between the two excitations as has been established by equation (\ref{noint}).\\
On the other hand, the spin magnitude defines the physical subspace
\begin{equation}
\{|n_a,n_b\rangle : n_a + n_b = N \}
\end{equation}
and the Schwinger transformation can be written as \cite{assa}
\begin{equation}\label{eq:jmas}
J_+=b^{\dagger}\sqrt{N-b^{\dagger}b},
\end{equation}
\begin{equation}\label{eq:jmenos}
J_-=\sqrt{N-b^{\dagger}b}\,\,b.
\end{equation}
The Hamiltonian of the bath in terms of the Schwinger bosons can be written as
\begin{equation}
H_B=\frac{g}{N}\left(b^{\dagger}b + a^{\dagger}a + 2a^{\dagger}a\left(b^{\dagger}b \right) \right)-g.
\end{equation}
Since $N=b^{\dagger}b + a^{\dagger}a$ and $a^{\dagger}a=N-N_b$, we can write the bath Hamiltonian as
\begin{equation}
H_B=g\left(1 + 2\left(1-\frac{N_b}{N} \right)b^{\dagger}b\right)-g.
\end{equation}
In the thermodynamic limit, $N\rightarrow\infty$, the Hamiltonian of the bath becomes
\begin{equation}
H_B= 2gb^{\dagger}b.
\end{equation}
The interaction Hamiltonian $H_{SB}$ given by equation (\ref{sb}) becomes
\begin{equation}
H_{SB}=\frac{g_0}{\sqrt{N}}\left(S^+a^{\dagger}b + S^-b^{\dagger}a\right)
\end{equation}
and using equations (\ref{eq:jmas}) and (\ref{eq:jmenos}) we obtain
\begin{equation}
H_{SB}=\frac{g_0}{\sqrt{N}}\left(S^+\sqrt{N-b^{\dagger}b}\,\,b + S^-b^{\dagger}\sqrt{N-b^{\dagger}b}\right)
\end{equation}
which in the thermodynamic limit, $N\rightarrow\infty$ gives
\begin{equation}
H_{SB}=g_0\left(S^+b+ S^-b^{\dagger}\right),
\end{equation}
and the total Hamiltonian can be written as
\begin{equation}\label{eq:jaynes}
H=\mu_0 S^z + 2gb^{\dagger} b + g_0(S^+b + S^-b^{\dagger}).
\end{equation}
This Hamiltonian is equivalent to the Hamiltonian of a two level atom in presence
of a bath of bosons of one single mode. It is clear that the Hamiltonian (\ref{eq:jaynes}) is analog to the
quantum model of radiation-matter interaction with a single mode in the rotating wave approximation: the Jaynes-Cummings model.

\noindent
The problem of spin wave theory can be studied by means of the Holstein-Primakoff transformation instead of Schwinger bosons. In this case the angular momentum operators are mapped to boson operators by \cite{primakoff}
\begin{equation}
J_+=b^{\dagger}\sqrt{2S-N_b} ,
\end{equation}
\begin{equation}
J_-=\sqrt{2S-N_b}\,\,b,
\end{equation}
and
\begin{equation}
J_z=N-N_a.
\end{equation}
It is easy to show that the operators $J$ obeys the angular momentum commutation relations,
\begin{equation}
[J_{\alpha},J_{\beta}]=i\epsilon^{\alpha\beta\gamma}J_{\gamma}.
\end{equation}
The two transformations known as Schwinger bosons, SB, and Holstein Primakoff, HP, are related by the following correspondence
\begin{center}
\begin{tabular}{ccl}
 SB & $\rightarrow$ &HP\\
	 b & $\rightarrow$ &b\\
	 a & $\rightarrow$ & $\sqrt{N-N_b}$.\\
\end{tabular}
\end{center}
As was discused before, this is a consequence of the physical subspace spanned. However, the Schwinger bosons transformation has a natural physical interpretation, as it was presented above.

\subsection{Reduced dynamics}
\noindent
Let us consider the solution of the dynamics of the density operator of the system to obtain the dynamics of the central spin. Suppose that the total system represented by $\rho(t)$ at $t=0$
is separable, $\rho(0)=|\psi\rangle \langle\psi|\otimes\rho_B $, where the initial state of the central spin is in a pure state, $|\psi\rangle$, and the bath is in a thermal bath given by the density operator $\rho_{Th}$:
\begin{equation}
\rho_B=\rho_{Th}=\frac{\exp(-H_B)/T}{Z},
\end{equation}
where the Boltzmann constant has been taken equal to one.
The Born approximation is made in the sense that the state of the bath is time independent, $\rho_B(t)=\rho_B$ for any time.
The partition function $Z$ is given by
\begin{equation}
Z=Tr[\exp(-H_B/T)],
\end{equation}
where the Hamiltonian of the bath, after the Schwinger transformation, is given by $H_B=2gb^\dagger b$. Then the partition function can be calculated analytically to obtain
\begin{equation}
Z=\frac{1}{1-e^{-2g/T}}.
\end{equation}
The dynamics of the total density operator of the system can be obtained from the Liouville-von Neumann equation,
\begin{equation}
\frac{d\rho}{dt}=-i[H,\rho],
\end{equation}
where $H$ is the total Hamiltonian given by the equation (\ref{eq:jaynes}).
This Hamiltonian can be written as
\begin{equation}
H=H_1+H_2
\end{equation}
where
\begin{equation}
H_1=2g b^{\dagger} b + 2g S^z
\end{equation}
\begin{equation}
H_2=\Delta S^z + g_0(S^+b + S^-b^{\dagger}).
\end{equation}
where $\Delta=\mu_o-2g$, usually known in quantum optics as \emph{detuning}. The two parts of this Hamiltonian commute, $[H_1,H_2]=0$, and the time evolution operator can be written as
\begin{equation}
U(t)=\exp(-iH t)=\exp(-iH_1t)\exp(-iH_2t)=U_1(t)U_2(t).
\end{equation}
In the two dimensional subspace spanned by the eigenvectors of $S^z$, that is in the base $\{|\uparrow\rangle,|\downarrow\rangle\}$, where  $|\uparrow\rangle$ and $|\downarrow\rangle$ indicate spin up and spin down respectively, the first factor, $U_1(t)$, of the time evolution operator is diagonal. On the other hand, the second factor, $U_2(t)$, is expanded in order to obtain,
\begin{equation}
U_2(t)=\left(
\begin{array}{cc}
\cos(g t\sqrt{b^{\dagger} b + 1}) & -i b\frac{\sin(gt\sqrt{b^{\dagger} b})}{\sqrt{b^{\dagger} b}}\\
-ib^{\dagger}\frac{\sin(g t\sqrt{b^{\dagger} b}+1)}{\sqrt{b^{\dagger} b+1}} & \cos(gt\sqrt{b^{\dagger} b})
\end{array}
\right),
\end{equation}
within the limit of zero detuning, $\mu_0=2g$, that is, with a fixed magnetic field of magnitude twice the magnitude of the internal interaction. In the two dimensional base,$\left\{\mid\uparrow \rangle, \mid \downarrow \rangle \right\}$, the evolution of the \emph{total} density operator is obtained by
\begin{equation}\label{eq:matriz}
\rho(t)=\hat{U}_2(t)\rho(0)\hat{U}^{\dagger}_2(t),
\end{equation}
and the central spin dynamics is obtained by tracing over the degrees of freedom of the environment, $\rho_{S}=Tr_B(\rho)$ \cite{cohen}.
The upper level population is given by
\begin{gather}\label{eq:population}
\rho_{uu}(t)=\sum_{n=0}^\infty P_n\cos^2(g_0\,t\sqrt{n+1}),
\end{gather}
while $\rho_{dd}=1-\rho_{uu}$, with $\rho_{uu}= \langle \uparrow |\rho_S(t)| \uparrow \rangle$ and $\rho_{dd}= \langle \downarrow |\rho_S(t)| \downarrow \rangle$.
The coefficients $P_n$, are the probability of the thermal mode $P_n=\langle n|\rho_{Th} | n \rangle$.
In this case  $P_n=\left(e^{-2gn/T}\right)\left(1-e^{-2g/T}\right)$. Note that the coefficients of the populations
$P_{n}$ include the effects of the interaction between the bath spins and the thermal energy $T$.
\begin{figure}[htb]
\centering
\includegraphics[width=7.5cm,height=5cm]{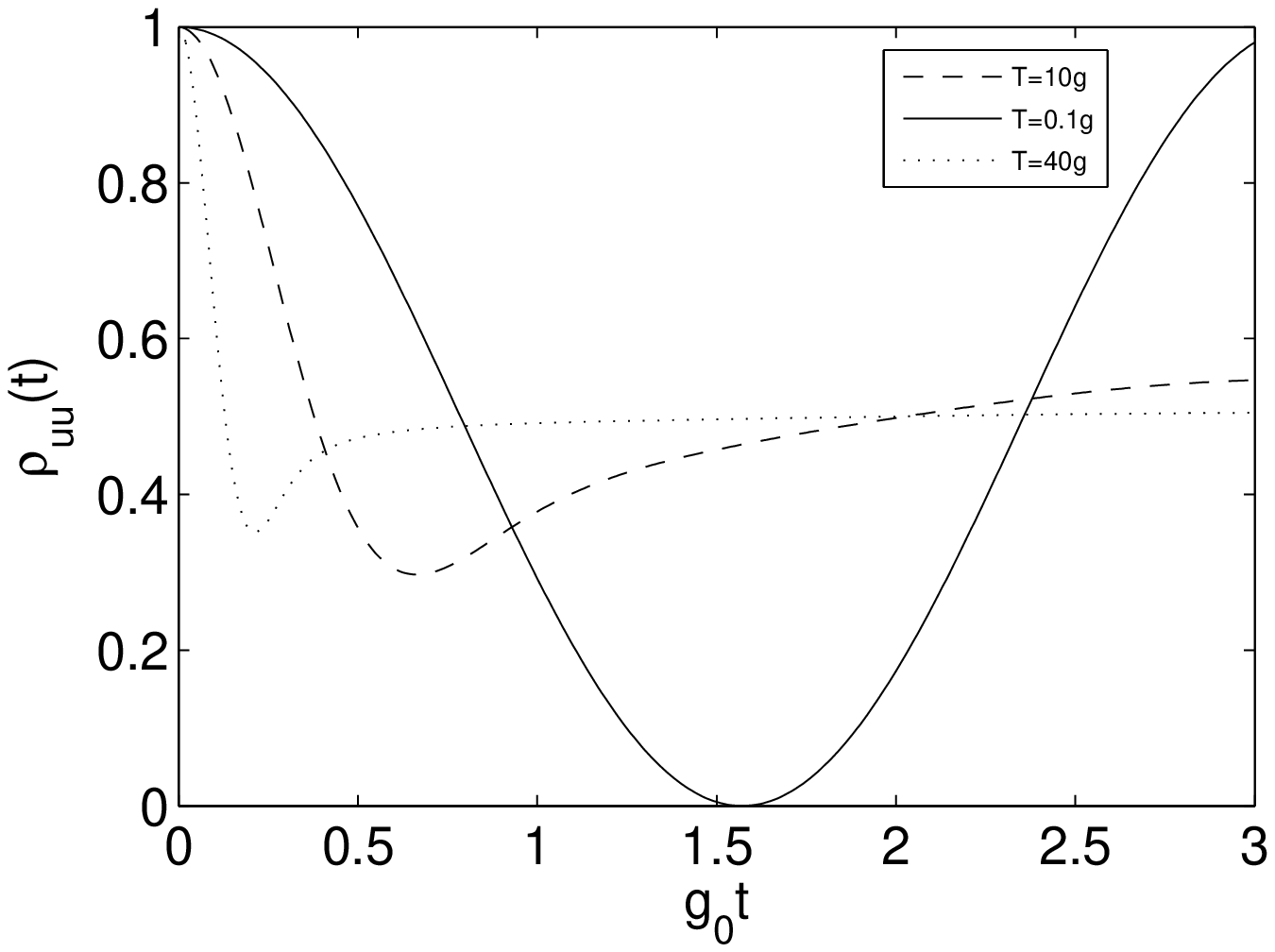}
\includegraphics[width=7.5cm,height=5cm]{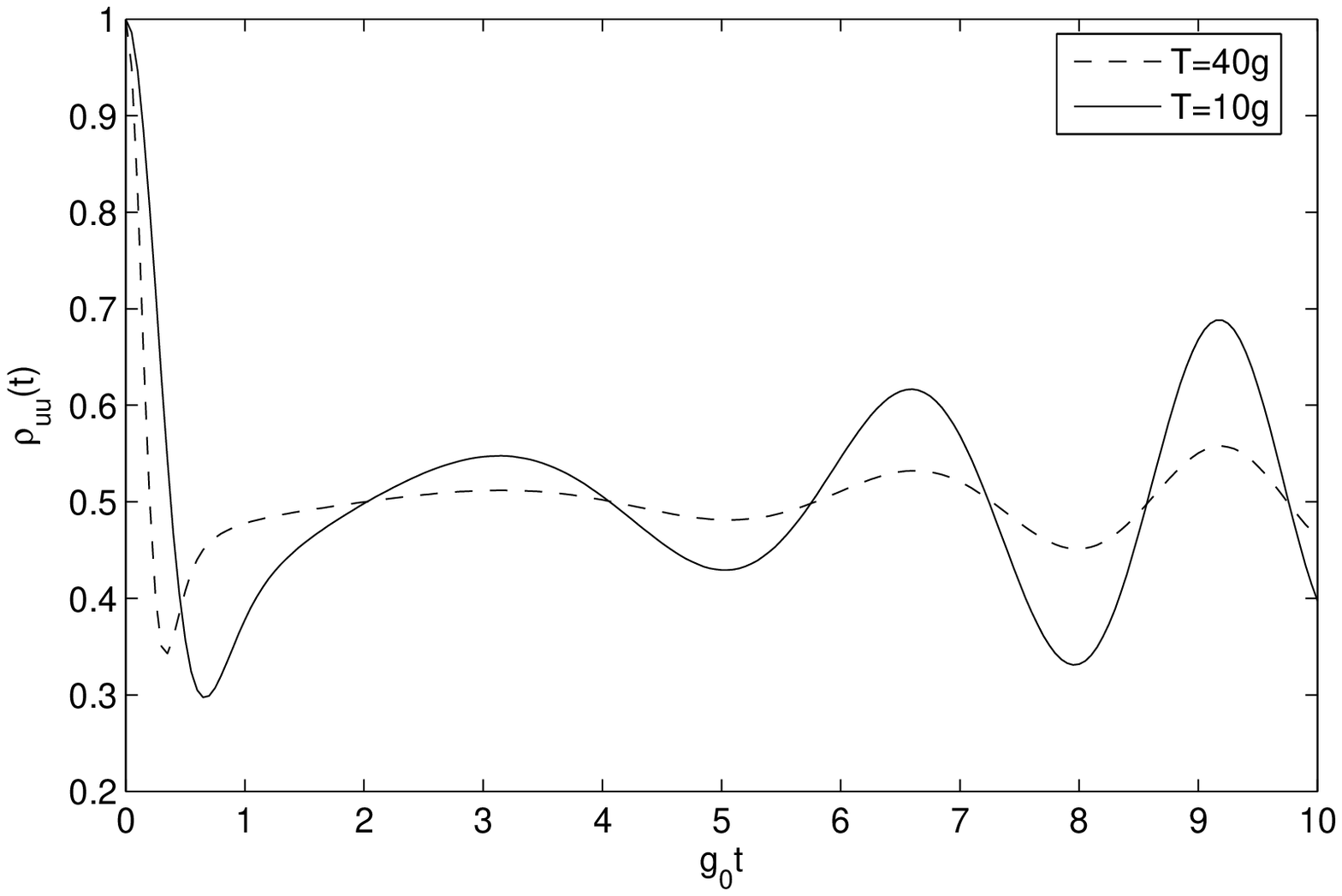}
\includegraphics[width=7.5cm,height=5cm]{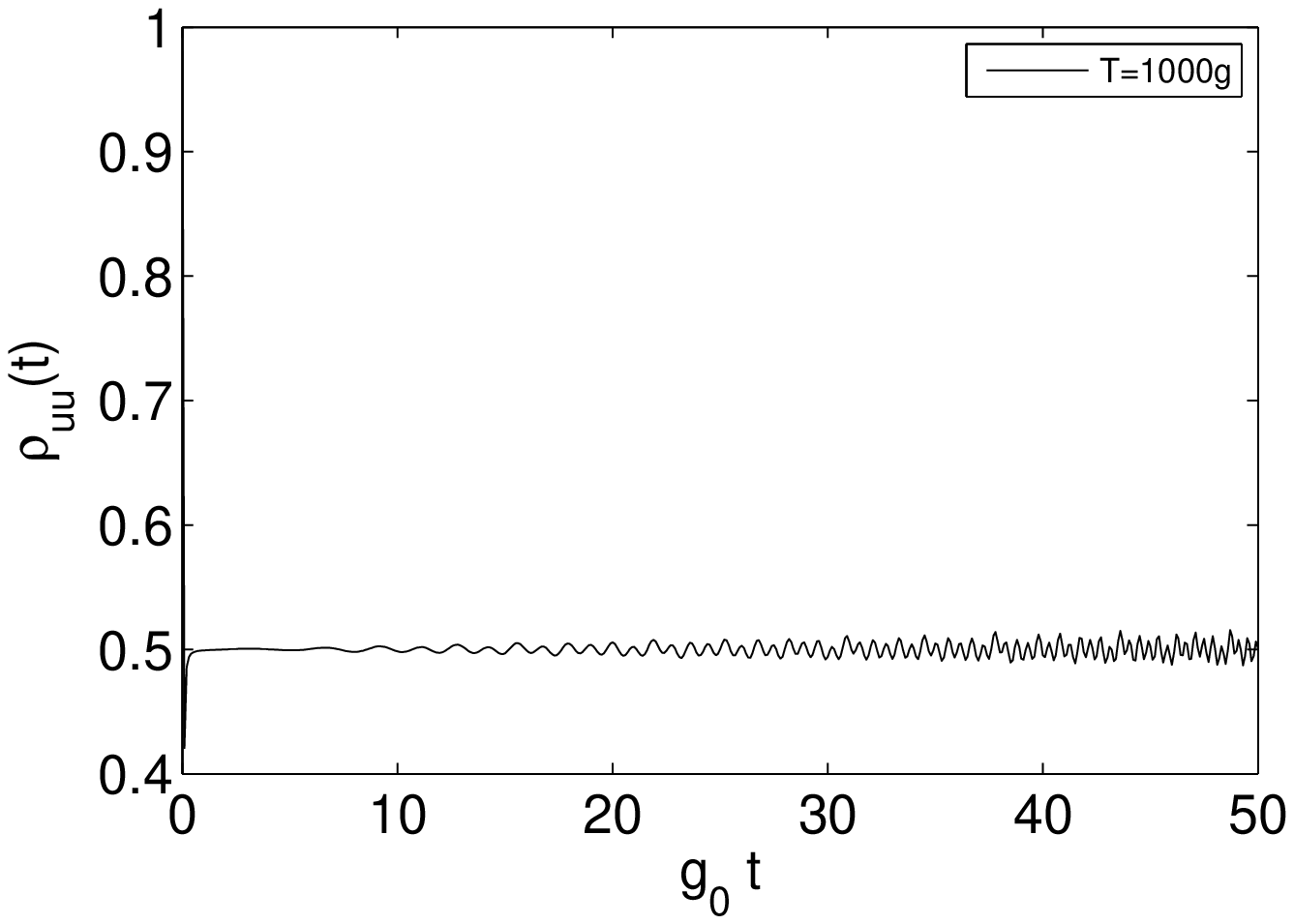}
\includegraphics[width=7.5cm,height=5cm]{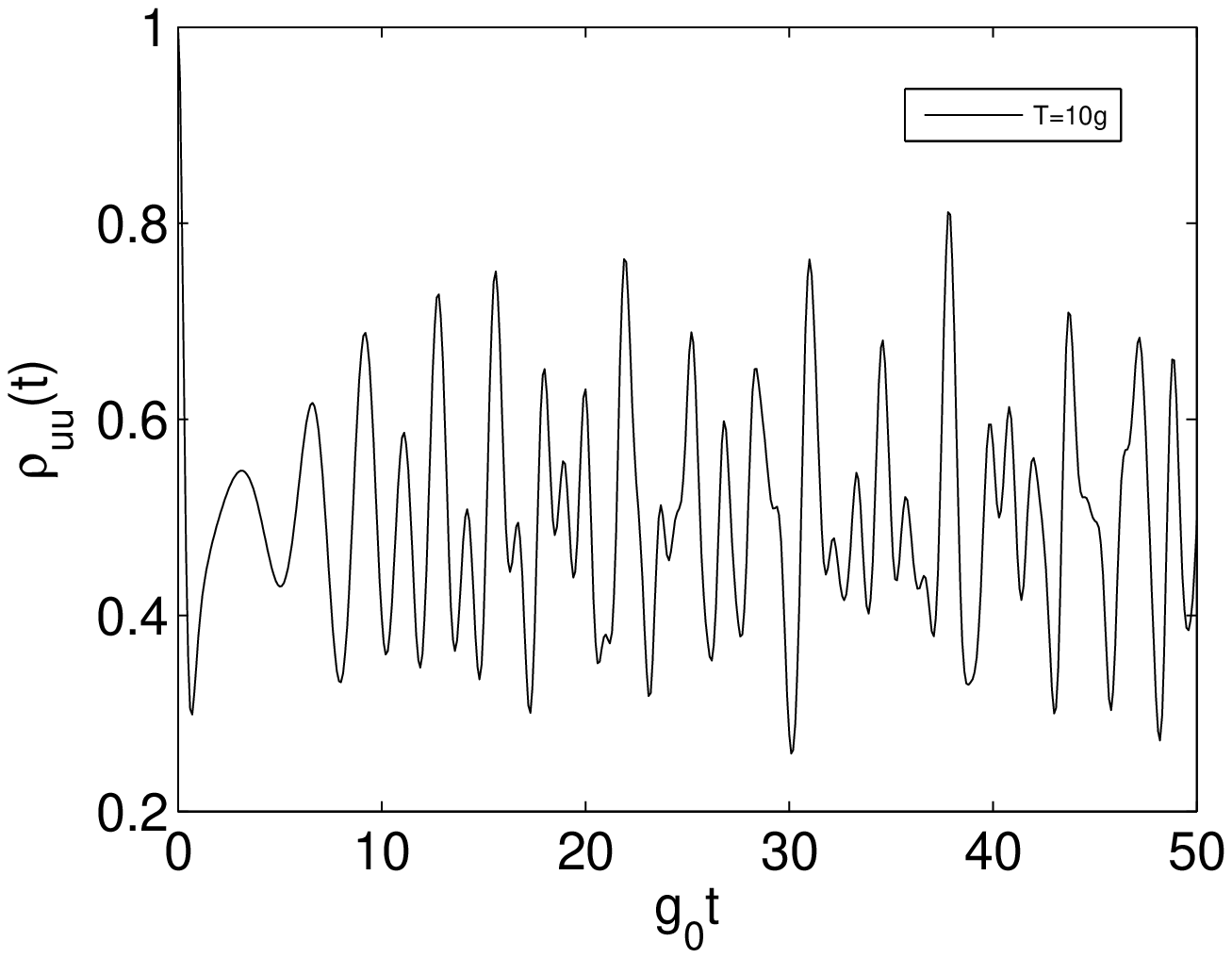}
\caption{The upper left plot shows the time evolution of the element $\rho_{uu}$ for three different temperatures $T=40g$, $T=10g$ and $T=0.1g$ with time in units of $g_0^{-1}$. The initial state of the central spin is $\mid \uparrow\rangle$. The upper right plot shows the
same result for a larger time scale and for the temperatures of $T=40g$ and $T=10g$. The down left plot shows the result obtained for a temperature of $T=1000g$, with a faster relaxation process. The lower right plot shows the graph but fora larger period of time and for a temperature of $T=10g$.  }\label{fig:prob}
\end{figure}
Figure \ref{fig:prob} shows the temporal evolution of the density matrix element, $\rho_{uu}$=$\langle\uparrow|\rho_S(t)|\uparrow\rangle$. The time is in units of $g_0^{-1}$ and the temperature in units of $g$. The initial state of the system is spin up $\mid\uparrow\rangle$. The effect of the temperature is clear in the sense that for increasing temperature  the relaxation process is faster. The plots show clear \emph{oscillations} of the relaxation process which get smaller as temperature increases. This is recognized as evidence of non Markovian dynamics (for example see reference \cite{Banyai}). The period of oscillation is approximately 3 units of $g_0^{-1}$ for almost all the temperatures and its amplitude decreases as the temperature increases, showing that the general effect of the temperature is to bring the system from pure states to mixed states.
Different temperatures have been considered. The upper left plot shows the time evolution of the element $\rho_{uu}$ for three different temperatures, $T=40g$, $T=10g$ and $T=0.1g$ with time in units of $g_0^{-1}$. The graph of the right shows the revivals of dynamics as a signal of the interchange of \emph{information} between  the system and the bath. The lower left plot shows the evolution of the population for a temperature of $T=1000g$. Although the temperature is high, in terms of the coupling constant $g$, it is possible to see that the memory effects still persist, but no so notable as in the case of low temperatures. The population oscillates around the value $0.5$, for which the system becomes a completely mixed state. The lower left graph shows that the oscillations never disappear as it is clear from the result for the upper level population, equation (\ref{eq:population}). 

\section{Von Neumann entropy dynamics}\label{sec:entropy}
\noindent
The von Neumann entropy  describes the departure of a system from a pure state or equivalently it measures the degree of mixture of a system. Taking two extreme values of zero for pure states and $\ln N$ for a maximally
 mixed state, N being the dimension of the Hilbert space.
The von Neumann entropy is defined as
\begin{equation}
S(\hat{\rho})\equiv-Tr[\hat{\rho}\ln \hat{\rho}],
\end{equation}
where $\hat{\rho}$ is the density matrix. An important property of the von Neumann entropy is that it 
is invariant under changes in the basis of $\rho$, and as we can see this property is primordial in order 
to calculate the evolution of the dunamics of the mixing.
\noindent    
Since the reduced density matrix is a $2\times2$ matrix, it can be written in terms of the Pauli matrices:
\begin{equation}\label{eq:bloch2}
    \hat{\rho}=
\frac{1}{2}\left(
\begin{array}{cc}
1+s_3    & s_1-is_2 \\
s_1+i s_2&1-s_3
\end{array}
\right)
    =\frac{1}{2}(\hat{I}+\mathbf{s}\cdot\sigma),
\end{equation}
where $\hat{I}$ is the identity matrix of dimension two, $\sigma$ are the \emph{traceless} Pauli matrices, and $\mathbf{s}$ is known as the Bloch vector.
The Bloch vector has magnitude one and its tip lies on the surface of the Bloch sphere when the state is pure, and it is easy to show that $\mid \mathbf{s} \mid\leq 1$ in the case of mixed state.
The eigenvalues $\lambda_1$ and $\lambda_2$ of the density matrix in terms of the components of the Bloch vector are:
\begin{equation}\label{eq:eigenvalues}
\begin{split}
    \lambda_1=\frac{1}{2}\big[1+\mid\hat{\mathbf{s}}\mid\big]\\
    \lambda_2=\frac{1}{2}\big[1-\mid\hat{\mathbf{s}}\mid\big].
\end{split}
\end{equation}
On the other hand, from the reduced density matrix (\ref{eq:matriz}), we can write it as
\begin{equation}\label{eq:reduced}
\hat{\rho}_S=
\left(
  \begin{array}{cc}
    \sum_{n=0}^{\infty} P_n\cos^2(g t\sqrt{n+1}) &  i\sum_{n=0}^{\infty} P_n\sin^2(g t\sqrt{n+1})\cos(g t\sqrt{n+1})    \\
    & \\
    & \\
    -i\sum_{n=0}^{\infty} P_n\sin^2(g t\sqrt{n+1})\cos(g t\sqrt{n+1}) & \sum_{n=0}^{\infty}P_n\sin^2(g t\sqrt{n+1}) \\
  \end{array}
\right)
\end{equation}

\noindent
where $P_n=e^{-2gn/T}(1-e^{2g/T})$. Comparing the matrix given by equation (\ref{eq:bloch2}) with the matrix given by equation (\ref{eq:reduced}) we obtain the elements of the Bloch vector, the eigenvalues of $\hat{\rho}$ given by the equation (\ref{eq:eigenvalues}), and the corresponding entropy of the system:
\begin{equation}\label{eq:entropy}
    S(t)=-\lambda_1\ln(\lambda_1)-\lambda_2\ln(\lambda_2).
\end{equation}
\begin{figure}[htb]
\centering
  \includegraphics[width=7.5cm, height=5cm]{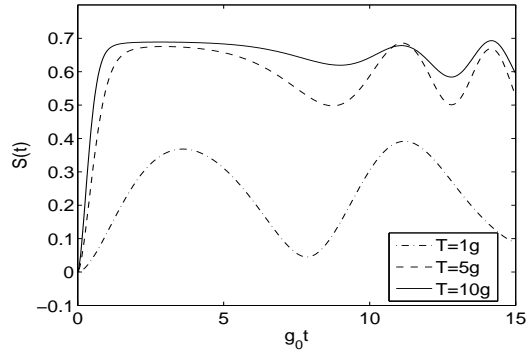}
\caption{Time evolution of the von Neumann entropy for the initial pure state $\mid \uparrow\rangle$ and for different values of the temperature, $T=1g$, $T=5g$ and $T=10g$. The larger the temperature, the faster that entropy becomes $S=\ln 2$: the entropy of a two level mixed state. The system and the bath becomes \emph{nearly} pure  for low temperatures.}\label{fig:entropia}
\end{figure}
\noindent
In figure \ref{fig:entropia} we present the plot of the von Neumann entropy corresponding to the reduced density matrix for the central spin in the case of a initial pure state $\mid \uparrow\rangle$. As in figure \ref{fig:prob}, the time is in units of $g_0^{-1}$, and the temperature in units of $g$. The result is notable in the sense that for low temperatures the central spin periodically returns to  its initial pure state. In the case of higher temperatures the system goes more rapidly to a completely mixed state of entropy  $S=\ln 2\simeq 0.69$, however, there remain oscillations with smaller amplitudes. In general, we can say that memory effects, or equivalently, non Markovian dynamics, are responsible of information gain for some particular time periods.
The result obtained for the entropy in the regime of high temperatures is consistent with \cite{breuer}, where the initial state was an infinite temperature unpolarized state. Some differences in the entropy of \cite{breuer} occur due to different initial states for the central spin.
\section{Conclusion}\label{sec:conclusion}
\noindent
The dynamics of a central spin in a bath of spins have been studied considering Heisenberg XY type interaction. An analytical solution was obtained for the reduced density matrix of the central spin in the thermodynamic limit.
The bath acts as bosonic collective excitations of one single mode via Schwinger bosons.
A separable state between the bath and the system was considered as initial state with the central spin in a pure state and the bath in a thermal state, evaluating the relaxation process for different temperatures. As the temperature increases, the relaxation process is faster as it is expected intuitively, however, memory or non Markovian effects are observed as oscillations of the populations which are reduced in amplitude with the increase of the temperature.
The evolution of the degree of mixure of the system in presence of a bath was studied by means of the von Neumann entropy for different temperatures. At low temperatures the system becomes periodically completely pure. The increase of the temperature implies only partial mixing and leads the system to a complete mixed state with entropy $S=\ln2$ in the case of infinity temperature. The results obtained coincide with that of the entropy of reference \cite{breuer}, where an unpolarized infinite temperature state was considered, but with the difference that in this case we can make calculations for different temperatures.
It is important to note that a thermodynamic limit was considered here, however, as was shown in \cite{breuer}, the results do not depend on the number of particles, $N$, for  $N>200$ for longer time scales; and for smaller times scales for $N\approx20$. We
point out the advantage of the thermodynamic limit permitting  analytical solution.

\noindent
We expect different behaviors of the entropy depending on the magnetic field as a control parameter of the degrre of mixing, as will be presented in a future work. On the other hand, our approach in terms of Schwinger bosons, permits to consider a general problem with clear physical interpretation giving also the possibility to describe a still more interesting system of interacting bath modes (work in progress).

\end{document}